\documentclass[a4paper,11pt]{article}

\usepackage{jcappub}

\usepackage{graphicx}
\usepackage{amsmath}
\usepackage{amssymb}
\usepackage{color}
\usepackage{subfig}

\newcommand\msun{\, \rm M_\odot}

\newcommand\kms{\, \rm km\,s^{-1}}

\newcommand\yr{{\, \rm yr}}
\newcommand\gyr{{\, \rm Gyr}}

\newcommand\eout{{e_{\rm out}}}
\newcommand\ein{{e_{\rm in}}}
\newcommand\aout{{a_{\rm out}}}
\newcommand\ain{{a_{\rm in}}}
\newcommand\mbh{{m_{\rm BH}}}
\newcommand\mns{{m_{\rm NS}}}
\newcommand\msmbh{{m_{\rm SMBH}}}
\newcommand\sigbh{{\sigma_{\rm BH}}}
\newcommand\signs{{\sigma_{\rm NS}}}
\newcommand\vk{{v_\mathrm{k}}}

\title{Supernovae in massive binaries and compact object mergers near supermassive black holes}
\author[a,b]{Giacomo Fragione}
\author[c]{Idan Ginsburg}
\author[c]{Abraham Loeb}

\affiliation[a]{Department of Physics \& Astronomy, Northwestern University, Evanston, IL 60202, USA}
\affiliation[b]{Center for Interdisciplinary Exploration \& Research in Astrophysics (CIERA), Evanston, IL 60202, USA}
\affiliation[c]{Astronomy Department, Harvard University, 60 Garden St., Cambridge, MA 02138, USA}

\emailAdd{giacomo.fragione@northwestern.edu}

\abstract{
Nuclear star clusters that surround supermassive black holes (SMBHs) in galactic nuclei are among the densest systems in the Universe, harbouring millions of stars and compact objects (COs). Within a few parsecs from the SMBH, stars can form binaries. In this paper, we model the supernova (SN) process of massive binaries that are born in proximity of the SMBH and that produce CO binaries. These binaries can later merge via emission of gravitational waves as a consequence of the Lidov-Kozai mechanism. We study the dynamical evolution of these systems by means of high-precision $N$-body simulations, including post-Newtonian (PN) terms up to 2.5PN order. We adopt different prescriptions for the natal velocity kicks imparted during the SN processes and find that larger kicks lead to more compact binaries that merge closer to the SMBH. We also conclude that most of the mergers enter the LIGO band with very high eccentricities. Finally, we compute a merger rate of $0.05$--$0.07\ \mathrm{Gpc}^{-3} \yr^{-1}$, $0.04$--$2\times 10^{-3}\ \mathrm{Gpc}^{-3} \yr^{-1}$, $9.6\times 10^{-6}$--$2.7\times 10^{-3}\ \mathrm{Gpc}^{-3} \yr^{-1}$ for BH-BH, BH-NS, NS-NS, respectively, smaller than the actual LIGO-Virgo observed rate.}

\begin{document}

\maketitle

\section{Introduction}

Ten black hole (BH) binaries and one neutron star (NS) binary mergers have been confirmed so far by the LIGO-Virgo collaboration\footnote{https://www.ligo.org - http://www.virgo-gw.eu} \cite{abbott16a,abbott16b,abbott17,abbott17b,abbott17c,ligo2018}, and a few new candidates have already been found in the O3 run. From these events, a merger rate of $9.7$--$101$ Gpc$^{-3}$ yr$^{-1}$ for binary BHs and a merger rate of $110$--$3840$ Gpc$^{-3}$ yr$^{-1}$ for binary NSs have been inferred. BH-NS mergers have not been observed yet and LIGO/Virgo has only set a $90\%$ upper limit of $610$ Gpc$^{-3}$ yr$^{-1}$ on the merger rate. The future observational campaigns of LIGO/Virgo and of the upcoming gravitational wave (GW) observatories are expected to find hundreds of signals, thus further constraining the CO merger rate and shedding light on their formation and evolution.

Several different astrophysical channels have been proposed to form merging compact objects, either through evolution in isolation or through dynamical interactions in dense environments, whose statistical weight can be disentangled based on the distributions of masses, spins, eccentricity and redshift \cite[see e.g.][]{olea16,gondan2018,fish2019}. Possible scenarios include isolated binary evolution \cite{bel16b}, binary chemically homogeneous evolution \cite{mand16,march16}, Lidov-Kozai (LK) mergers of binaries in galactic nuclei \cite{antoper12,petr17,fragrish2018,grish18}, in isolated triples \cite{ant17,sil17,arc2018} and quadruples \cite{fragk2019,liu2019}, mergers in globular and nuclear star clusters \cite{askar17,baner18,frak18,rod18} and mergers in accretion disks of active galactic nuclei \cite{bart17}.

Our Galactic Centre (GC) contains a large population of young massive O-type stars, many of which have been observed to reside in a stellar disk and probably were born in-situ \cite{genz10}. Three binaries have been observed within $\sim 0.1$ pc of the GC \cite{ott1999,pfu2014}, and S0-27 has been suggested to be the first potential binary in the S-star population \cite{jia2019}. Some of the binaries may be due to collisions and subsequent mergers \cite{gins2007}. These massive stars can evolve and eventually form compact object (CO) binaries \cite{prod2015,steph2016M,bort2017,fragione19,steph2019}, which form a hierarchical three-body system with the supermassive black hole (SMBH). As a result, they can undergo inclination and eccentricity oscillations due to the LK mechanism \cite[see][for a review]{naoz2016}. If the excursion in eccentricity is significant, the binary would dissipate energy near pericentre, shrink its orbit, and eventually merge much faster than in isolation \cite{antoper12,fragrish2018,hamers18,hoang18}.

In this paper, for the first time, we model the supernova (SN) processes in massive binaries that give birth to a BH or a NS in the proximity of an SMBH, and study the following dynamical evolution of BH and NS binaries by means of high-precision $N$-body simulations, including PN terms up to 2.5PN order. We start from the main-sequence (MS) progenitors of BHs and NSs and model the SN events that lead to the formation of these COs. We denote the BH and NS masses as $\mbh$ and $\mns$, respectively, and the mass of the SMBH as $\msmbh$. We adopt a variety of prescriptions for the natal kick velocities that are imparted by SN events and quantify how the merger rates depend on the initial conditions, which we compare to other GW merger scenarios.

The paper is organized as follows. In Section~\ref{sect:nuclei} we discuss the relevant timescales and the properties of the BHs and NSs in galactic nuclei. In Section~\ref{sect:initialcond} we present our numerical methods, while in Section~\ref{sect:results} we discuss the results and determine the rate of BH and NS mergers in the vicinity of an SMBH. Finally, in Section~\ref{sect:conc} we discuss the implications of our findings and draw our conclusions.

\section{Galactic nuclei}
\label{sect:nuclei}

\subsection{Timescales}

We start by summarizing the relevant processes that take place in the dense stellar environment of galactic nuclei. We consider a triple that consists of a binary of total mass $m_b=m_1+m_2$ orbiting a SMBH of mass $\msmbh$. The semi-major axis and eccentricity of the binary inner orbit are $\ain$ and $\ein$, respectively, while the semi-major axis and eccentricity of the outer orbit are $\aout$ and $\eout$, respectively. The initial mutual inclination of the inner and outer orbital planes is $i_0$.

Within its radius of influence $r_h$, the potential of the SMBH dominates the dynamics of stars and COs, which describe near-Keplerian orbits \cite{mer13}. Stars and COs diffuse in energy and angular momentum, as a result of continuous non-coherent scattering events with other stars and COs. This process takes place over a 2-body timescale \cite{binntrem87},
\begin{eqnarray}
T_{2B}&=&1.6\times 10^{10}\mathrm{yr}\ \left(\frac{\sigma}{300\ \mathrm{km s}^{-1}}\right)^3\left(\frac{m}{\msun}\right)^{-1}\times\nonumber\\
&\times&\left(\frac{\rho}{2.1\times 10^{6}\msun\ \mathrm{pc}^{-3}}\right)^{-1}\left(\frac{\ln \Lambda}{15}\right)^{-1} \ .
\label{eqn:t2b}
\end{eqnarray}
where $\rho$ and $\sigma$ are the 1D density and velocity dispersion, respectively, $\ln \Lambda$ is the Coulomb logarithm and $m$ is the average stellar mass. Typically, $T_{2B}\gtrsim 10^9\yr$. \cite{bahcall76} showed that a population of equal-mass objects forms a power-law density cusp around a SMBH, $n(r)\propto r^{-\alpha}$, where $\alpha=7/4$. When a mass spectrum is taken into account, lighter and heavier objects develop shallower and steeper cusps, respectively \cite{ale09}, as a result of the dynamical friction acting on the more massive bodies. Source terms tend to make the cusp steeper as well \cite{aharon15,frasar18}. Only galactic nuclei that host an SMBH less massive than $\sim 10^7\ \mathrm{M}_{\odot}$ have typical 2-body timescales short enough to make the effects of the non-coherent stellar interactions important within a Hubble time.

On timescales smaller than $T_{\rm 2B}$, the residual torque becomes relevant and the angular momentum diffuses both in direction and magnitude \cite{rauch96}. This process takes place over a resonant relaxation timescale \cite{kocs15},
\begin{equation}
T_{\rm RR}=9.2\times 10^{8}\ \mathrm{yr}\ \left(\frac{\msmbh}{4\times 10^6 \msun}\right)^{1/2}\left(\frac{a_{out}}{0.1\ \mathrm{pc}}\right)^{3/2}\left(\frac{m}{\msun}\right)^{-1}\ .
\label{eqn:trr}
\end{equation}
Typically, $T_{\rm RR}\sim 10^7\yr$--$10^{10}\yr$.

On shorter timescales, vector resonant relaxation changes the direction of the orbital angular momentum on a typical timescale \cite{kocs11,kocs15},
\begin{eqnarray}
T_{\rm VRR}&=&7.6\times 10^{6}\ \mathrm{yr}\ \left(\frac{\msmbh}{4\times 10^6 \msun}\right)^{1/2}\times\nonumber\\
&\times & \left(\frac{a_{\rm out}}{0.1\ \mathrm{pc}}\right)^{3/2}\left(\frac{m}{\msun}\right)^{-1}\left(\frac{N}{6000}\right)^{-1/2}\ .
\label{eqn:tvrr}
\end{eqnarray}
Here $N$ is the number of stars within $a_{\rm out}$. Typically, $T_{\rm VRR}\sim 10^5\yr$--$10^7\yr$.

Binaries may evaporate due to dynamical interactions with stars and COs in the dense environment of a galactic nucleus. This process takes place on an evaporation timescale \cite{binntrem87},
\begin{eqnarray}
T_{\rm EV}&=&3.2\times 10^{7}\ \mathrm{yr} \left(\frac{m_1+m_2}{2\msun}\right)\left(\frac{\sigma}{300\ \mathrm{km s}^{-1}}\right)\left(\frac{\msun}{m}\right)\nonumber\\
&\times&\left(\frac{a_{\rm in}}{1\ \mathrm{AU}}\right)^{-1}\left(\frac{2.1\times 10^{6}\msun\ \mathrm{pc}^{-3}}{\rho}\right)\left(\frac{15}{\ln \Lambda}\right)\ .
\label{eqn:binevap}
\end{eqnarray}
If binaries do not evaporate and the initial mutual orbital inclination of the inner and outer orbit is in the range $\sim 40^\circ$-$140^\circ$, the binary experiences eccentricity oscillations on a LK timescale \cite{lid62,koz62},
\begin{equation}
T_{\rm LK}=\frac{8}{15\pi}\frac{m_{\rm tot}}{m_{\rm SMBH}}\frac{P_{\rm out}^2}{P_{\rm in}}\left(1-e_{\rm out}^2\right)^{3/2}\ .
\end{equation}
In the previous equation, $P_{\rm bin}$ and $P_{\rm out}$ are the binary period and its orbital period with respect to the SMBH, respectively, and $m_{\rm tot}=m_1+m_2+\msmbh\approx\msmbh$. At the quadrupole order, the eccentricity can approach unity if $i_0\sim 90^\circ$. At the octupole order of approximation, $\ein$ can reach almost unity even if the initial inclination is outside of the LK window angle, if the outer orbit is eccentric \cite{naoz13a}. This happens over the octupole timescale,
\begin{equation}
T_{\rm oct}=\frac{1}{\epsilon}T_{\rm LK}\ ,
\label{eqn:tlkoct}
\end{equation}
where the octupole parameter is defined as,
\begin{equation}
\epsilon=\frac{m_1-m_2}{m_1+m_2}\frac{\ain}{\aout}\frac{\eout}{1-e_{\rm out}^2}\ .
\end{equation}
However, general relativistic precession might suppress LK oscillations \cite{li2017}.

\subsection{Compact objects}

Next, we address the current theoretical and observational understanding of COs in galactic nuclei. Observationally, only our own galactic nucleus is sufficiently close to resolve single binaries containing COs. 

\cite{hailey17} reported observations of a dozen quiescent X-ray binaries that form a central density cusp within $\sim 1$ pc from the SMBH in our Galactic Centre. Six more X-ray transients are known to be present in the inner parsec of our Galaxy \cite{muno05,hailey17}. \cite{generosoz18} combined the reported statistics from the literature and estimated that the number of NS X-ray binaries per stellar mass in the Galactic Centre is comparable to the number expected to be in globular clusters and about three orders of magnitude higher than in the field. Likewise, the number of BH X-ray binaries per stellar mass is $\sim 3$ orders of magnitude higher than in the field, and about an order of magnitude higher than in globular clusters \cite{strader12,leigh14,leigh16}. 

These large numbers of NS and BH binaries must originate from specific channels. Yet, little is known about binaries in our Galactic Centre and other galactic nuclei. A likely origin for the BHs and NSs is as remnants of massive O/B stars\footnote{BHs may originate from the merger of two NSs, while NSs from the merger from two WDs.}. Our Galactic Centre contains a large population of young massive O-type stars, many of which have been observed to reside in a stellar disk and were likely born in-situ within the central $\sim 0.5$ pc \cite{genz10}. These stars are observed to be
relatively more massive and to follow a top-heavy initial mass function \cite{lu2013}, probably as a consequence of the higher pressure of the interstellar gas (and hence higher Jeans mass) typical of galactic nuclei. In the Solar neighborhood, the binary fraction is $\gtrsim 70\%$ for young massive stars \cite{sana2012}. If a similar fraction holds for galactic nuclei, this would suggest a high rate of BH and NS binary formation. Eventually, binaries comprising NSs and BHs can form outside the nuclear cluster and then migrate inwards on a 2-body timescale \cite{antoper12}, or can be delivered near the SMBH as a result of triple/quadruple disruptions \cite{per09,fgu18,fgi18} and infalls of star clusters \cite*{antmer,fck17}. 

Whatever their origin is, BH and NS distributions are shaped by the dynamical processes described above. Since BHs and NSs are heavier than average stars, they would relax into steeper cusps, with BHs relaxing into steeper cusps than NSs \cite{Fre06,hopal06,emam2019}. In case BHs have a mass spectrum, only the more massive ones would follow a steeper distribution \cite{aharon16}. In case BHs and NSs form binaries, the typical distribution of semi-major axes is not well known, in part due to the rich dynamics in the dense environment of galactic nuclei which causes the distribution to diffuse over time. Regarding eccentricity, an \textit{in-situ} formation would probably favour circular binaries, while the migration scenario would rather prefer a thermal distribution, as a consequence of the energy exchange through many dynamical encounters \cite{geller2019}. However, the properties of stellar and CO binaries may differ from what we observe in the Solar neighborhood and in isolation, since both the large tidal field of the nearby SMBH and the abundant stars and COs shape the semi-major axis and eccentricities of these binaries \cite{hop09}.

\begin{table*}
\caption{Models parameters: name, slope of the initial mass function, CO binary, dispersion of BH kick-velocity distribution ($\sigbh$), dispersion of the NS kick-velocity distribution ($\signs$), eccentricity distribution ($f(e)$), fraction of stable systems after SNe ($f_{\rm stable}$), fraction of mergers from the $N$-body simulations ($f_{\rm merge}$).}
\centering
\begin{tabular}{lccccccccc}
\hline
Name & $\beta$ & type & $\sigbh$ ($\kms$) & $\signs$ ($\kms$) & $f(e)$ & $f_{\rm stable}$ & $f_{\rm merge}$\\
\hline\hline
A1 & $2.3$ & BH-BH & $\signs\times m_\mathrm{NS}/m_\mathrm{BH}$ & $260$ & uniform  & $1.2\times 10^{-1}$ & $0.026$\\
A2 & $2.3$ & BH-BH & $0$  & $0$ & uniform  & $1.7\times 10^{-1}$ & $0.023$\\
A3 & $1.7$ & BH-BH & $\signs\times m_\mathrm{NS}/m_\mathrm{BH}$ & $260$ & uniform  & $1.4\times 10^{-1}$ & $0.037$\\
B1 & $2.3$ & BH-NS & $\signs\times m_\mathrm{NS}/m_\mathrm{BH}$ & $260$ & uniform  & $9.4\times 10^{-2}$ & $0.027$\\
B2 & $2.3$ & BH-NS & $0$  & $0$ & uniform  & $4.4\times 10^{-3}$ & $0.023$\\
C1 & $2.3$ & NS-NS & $\signs\times m_\mathrm{NS}/m_\mathrm{BH}$ & $260$ & uniform  & $2.4\times 10^{-5}$ & $0.017$\\
C2 & $2.3$ & NS-NS & $0$  & $0$ & uniform  & $6.7\times 10^{-3}$ & $0.010$\\
\hline
\end{tabular}
\label{tab:models}
\end{table*}

\section{Initial conditions}
\label{sect:initialcond}

The stellar triples in our simulations are initialized as follows. In total, we consider six different models (see Table~\ref{tab:models}) for a Milky Way-like nucleus, where we set $\msmbh=4\times 10^6\msun$.

For simplicity, we assume that every star in the mass range $8 \msun$--$20\msun$ will collapse to a NS of mass $\mns=1.3\msun$, while a star in the mass range $20 \msun$--$150\msun$ forms a BH of mass $\mbh=m/3$. We note that the exact values of $\mns$ and $\mbh$ can depend on the details of stellar evolution related to metallicity, rotation and stellar winds. For example, lower-metallicity progenitors can collapse to make a heavier BH then higher-metallicity progenitors \cite[e.g.][]{spera2015}. Stellar winds could also be important since they may change the orbital parameters of binaries before a BH or a NS forms. The picture is even more complicated if mass loss during possible episodes of Roche-lobe overflows and common envelope evolution phases are taken into account. These processes are highly uncertain even in binaries, and even more complicate in triple systems because of the interplay with the LK cycles imposed by the tertiary \cite{rosa2019,hamd2019}. Therefore, we prefer the above simple procedure to derive the final BH and NS masses, which nevertheless gives a maximum BH mass consistent with recent theoretical results on pulsational pair instabilities that limit the maximum BH mass to $\sim 50\msun$ \cite{bel2016b}. In all our models, we sample the mass $m_1$ of the most massive star in the inner binary from an initial mass function,
\begin{equation}
\frac{dN}{dm} \propto m^{-\beta}\ ,
\label{eqn:bhmassfunc}
\end{equation}
in the range $8 \msun$-$150 \msun$. We set $\beta=2.3$ as for a canonical \cite{kroupa2001} mass function. We run an additional model where we take $\beta=1.7$ for CO that originated from stellar binaries in stellar disks, which are observed to be
relatively more massive and follow a top-heavy initial mass function \cite{lu2013}. We adopt a flat mass ratio distribution for the inner binary, $m_2/m_1$, consistent with observations of massive binary stars \cite{sana2012,duch2013,sana2017}. We consider three different types of binaries that form this way: BH-BH, BH-NS, NS-NS. We assume that the distribution of $\ain$ to be flat in log-space, roughly consistent with the results of \cite{kob2014}. We set as minimum and maximum separation $1$ AU and $50$ AU, consistent with Sana's distribution, which favors short period binaries \cite{sana2012}. For the outer semi-major axis $\aout$, we assume that the CO binaries follow a power-law density cusp around the SMBH, $n(r)\propto r^{-\alpha}$, where $\alpha=7/4$. For the orbital eccentricities of the inner binary, $\ein$, and outer binary, $\eout$, we assume a flat distribution. Finally, the initial mutual inclination $i_0$ between the inner and outer orbit is drawn from an isotropic distribution (uniform in $\cos i_0$), while the other relevant angles are drawn randomly. After sampling the relevant parameters, we check that the initial configuration satisfies the stability criterion of hierarchical triples of Mardling \& Aarseth \cite{mar01}, otherwise we sample again the relevant parameters.

Given the above set of initial conditions, we let the primary star and the secondary star in the massive binary to undergo a conversion to a BH or a NS, sequentially. This means that we let the primary undergo a SN event, then we update the orbital parameters of the triple system (binary+SMBH) as in \cite{fragl19}. If it is bound and stable, we let the secondary undergo a SN event. We update again the orbital parameters of the system and, if the system is still stable and bound, we integrate the newly formed CO binary to determine if it leads to a GW event. As a result of the mass loss from the primary and secondary star, the binary is imparted a kick to its center of mass \cite{bla1961}. Furthermore, the system receives a natal kick due to recoil from an asymmetric supernova explosion. We assume that the natal velocity kick is drawn from a Maxwellian distribution,
\begin{equation}
p(\vk)\propto \vk^2 e^{-\vk^2/\sigma^2}\ ,
\label{eqn:vkick}
\end{equation}
with a mean velocity $\sigma$. The distribution of natal kick velocities of BHs and NSs is highly uncertain. \cite{hobbs2005} derived a distribution which can be approximated by a Maxwellian distribution with dispersion $\sim 260\kms$ for NSs, while \cite{arz2002} inferred a double-peaked distribution with a peak at $\sim 100\kms$ and a smaller peak at $\sim 700\kms$ for NSs. We implement momentum-conserving kicks, in which we assume that the momentum imparted to a BH is the same as the momentum given to a NS \cite{fryer2001}. As a consequence, the kick velocities for the BHs will be reduced with respect to those of NSs by a factor of $\mbh/\mns>1$. We adopt two different values of the velocity dispersion $\sigma$, to sample the possible range of this parameter. In some models, we consider a non-zero natal kick velocity for the newly formed BHs and NSs by adopting $\sigma=\signs=260 \kms$. We also run a model where no natal kick is imparted during the CO formation (i.e. $\sigma=0\kms$). For NSs, this would be consistent with the electron-capture supernova process, where the newly-formed NSs would receive no kick at birth or only a very small one due to an asymmetry in neutrino emissions \citep{pod2004}. We note that even in this case, the system experiences a kick to its center of mass because one of the massive components suddenly loses mass \citep{bla1961}.

After each SN event, the orbital elements of the triple are updated as appropriate \cite{pijloo2012,lu2019,fragl19}, to account both for mass loss and natal kicks \cite{bla1961}. We also check again that the stability criterion of hierarchical triples of Mardling \& Aarseth \cite{mar01} is satisfied and the system is stable. After all the SNe take place, we integrate the triple systems by means of the \textsc{ARCHAIN} code \cite{mik06,mik08}, a fully regularized code able to model the evolution of binaries of arbitrary mass ratios and eccentricities with high accuracy and that includes PN corrections up to order PN2.5. We perform $\sim 750-1000$ simulations for each model in Table~\ref{tab:models}. We fix the maximum integration time to be,
\begin{equation}
T=\min \left(T_{\rm EV}, 10\ \gyr \right)\ ,
\label{eqn:tint}
\end{equation}
where $T_{\rm EV}$ is the evaporation time of Eq.~\ref{eqn:binevap} computed for the newborn CO binary. As a consequence of the SNe, the evaporation time of binaries becomes shorter on average. We illustrate this in Fig.~\ref{fig:tev}, where we show the evaporation time of BH-BH and NS-NS binaries in a Milky Way-like galaxy pre-SNe and post-SNe, for $\sigma=0\kms$ and $\sigma=260\kms$. To compute $T_{\rm EV}$, we have assumed the stars are distributed according to a Bahcall-Wolf cusp \cite{bahcall76}. We find that $\sim 10\%$, $\sim 15\%$, $\sim 40\%$ the systems that lead to a BH-BH, BH-NS and NS-NS, respectively, have pre-SN $T_{\rm EV}$ shorter than the evolutionary timescale of each star in the binary. These systems would eventually evaporate before collapsing to a CO binary as a result of dynamical interaction with other stars and COs.

\section{Results}
\label{sect:results}

\subsection{Inclination distribution}

\begin{figure} 
\centering
\includegraphics[scale=0.525]{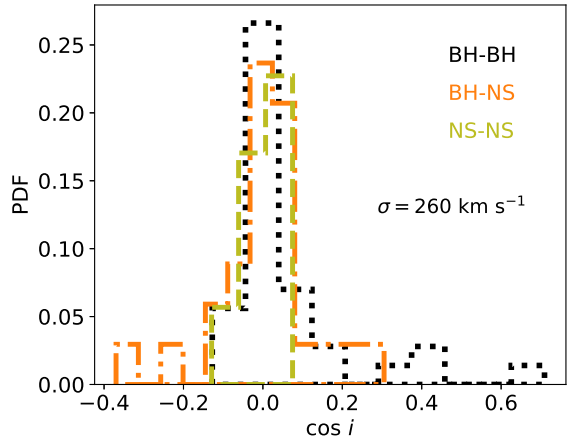}
\caption{Inclination ($i$) distribution of BH-BH, BH-NS and NS-NS binaries that merge in a Milky Way-like galaxy. Most of the systems merge when $i\sim 90^\circ$, where the effect of the LK oscillations is maximal.}
\label{fig:incl}
\end{figure}

CO binaries are expected to be significantly perturbed by the tidal field of the SMBH whenever its orbital plane is sufficiently inclined, i.e. $i\sim 90^\circ$, with respect to the outer orbit \cite{lid62,koz62}. Figure~\ref{fig:incl} depicts the probability distribution function (PDF) of $i$ in the systems which produce a merger. The distributions are shown for BH-BH, BH-NS, and NS-NS binaries in the case $\sigma=260\kms$. The majority of the CO mergers happen when the inclination is $\sim 90^\circ$, consistent with the quadrupole LK theory. In this case, the LK cycles can lead the CO eccentricity up to unity and binaries can efficiently dissipate energy through gravitational radiation at pericentre, eventually producing a GW merger event.

\subsection{Semi-major axis distribution}

\begin{figure}
\centering
\includegraphics[scale=0.525]{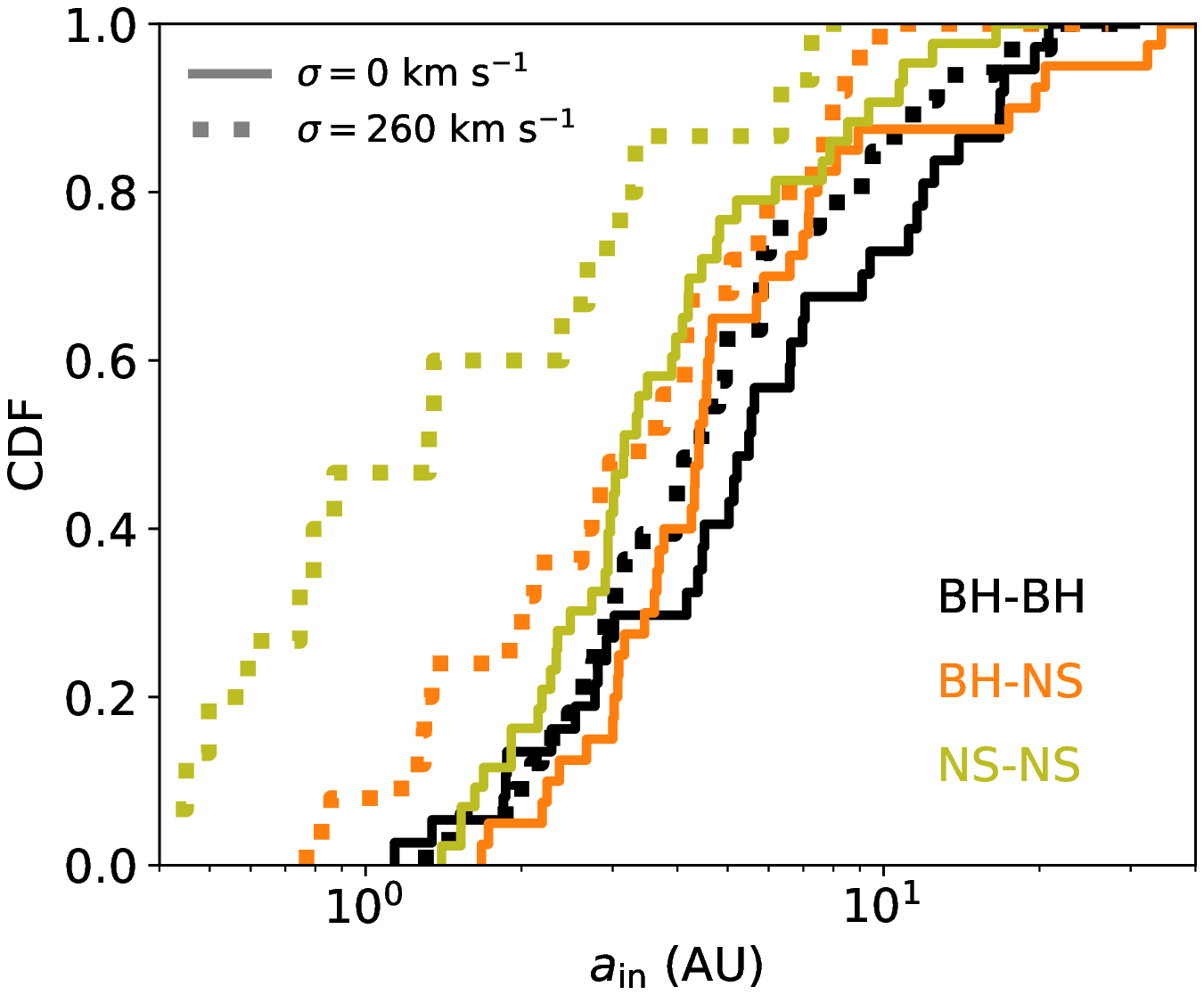}
\includegraphics[scale=0.525]{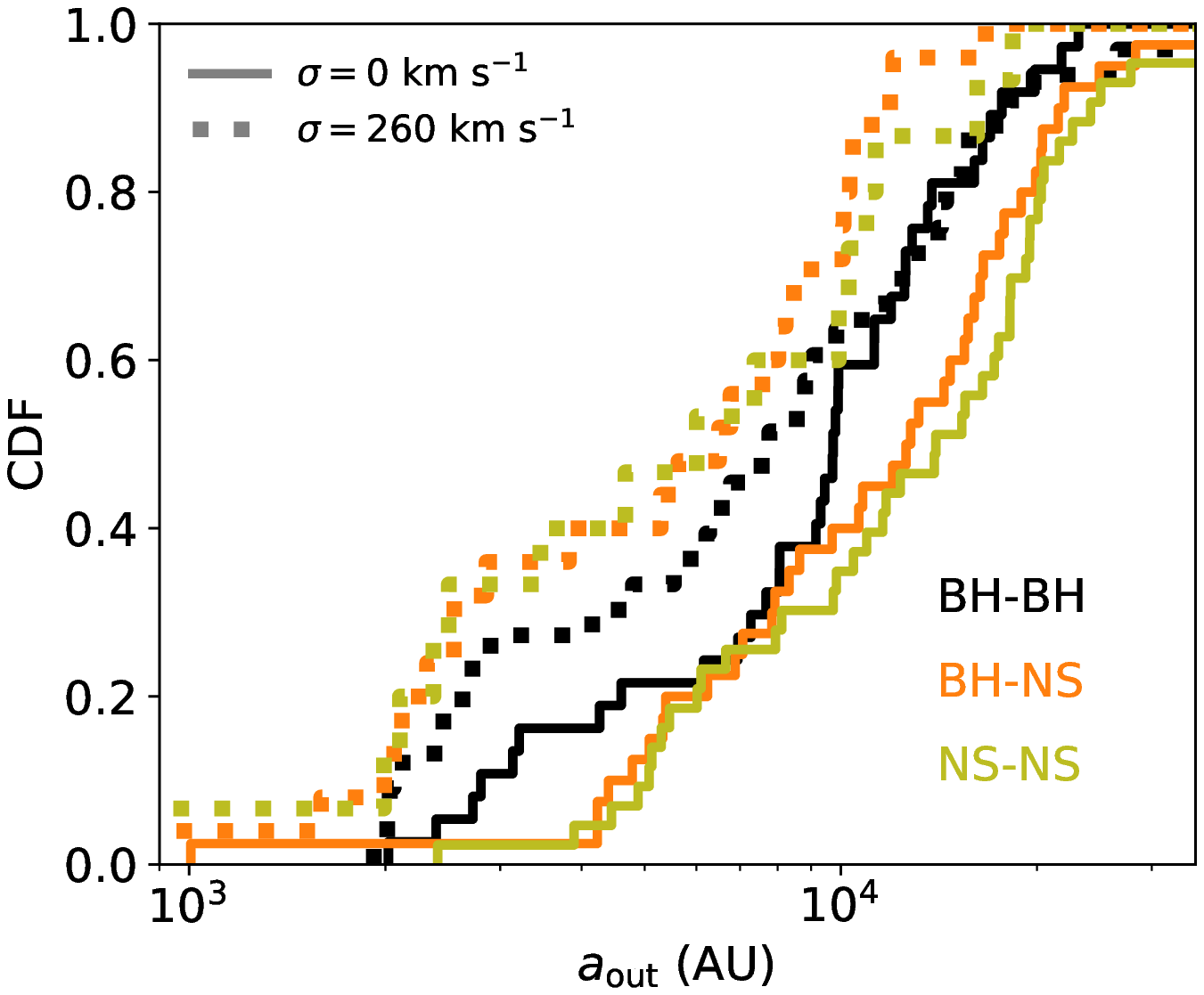}
\caption{Cumulative distribution function of inner (left) and outer (right) semi-major axis of binaries that lead to a merger, for different values of $\sigma$.}
\label{fig:semi}
\end{figure}

Figure~\ref{fig:semi} shows the cumulative distribution function (CDF) of inner (top) and outer (bottom) semi-major axes of BH-BH, BH-NS, and NS-NS binaries that lead to a merger for $\sigma=0\kms$ and $\sigma=260\kms$. The $a_{\rm in}$ and $a_{\rm out}$ we report are the inner and outer semi-major axes after the SN events in both stars take place. Merging binaries have typically smaller inner and outer semi-major axes in the case the kick-velocity is drawn from a Maxwellian with a larger $\sigma$. The main motivation is that systems with initially wide orbits are generally unbound by large kick velocities. If they stay bound, they could however become so wide to evaporate on a short $T_{\rm EV}$ or to be unstable according to the Mardling \& Aarseth criterion \cite{mar01}. Moreover, we note that SN kicks can result in binaries with much a much smaller semi-major axis than pre-explosion, in the case the kick is opposite to the orbital velocity. This can explain the fraction of systems with $\ain\lesssim 1$ AU for the NS-NS population. In the case of Blaauw kicks, the semi-major axis always increases, thus justifying the populations with $\ain\gtrsim 1$ AU for the case of $\sigma=0\kms$. We find that $\sim 50\%$ of the BH-BH, BH-NS, and NS-NS mergers have $\ain\lesssim 6$ AU, $\lesssim 5$ AU, and $\lesssim 3$ AU for $\sigma=0\kms$ and $\ain\lesssim 4$ AU, $\lesssim 3$ AU, and $\lesssim 1$ AU for $\sigma=260\kms$. As shown in Figure~\ref{fig:semi}, CO binaries that merge closer to the SMBH are expected to be produced after a larger SN kick is imparted to the exploding star. We note that, however, wider CO binaries orbiting an SMBH evaporate on shorter timescales as a result of the dynamical interactions with other stars and COs, which are more frequent closer to the SMBH, before the LK cycles can drive them to merger. As a consequence, the shift in the distributions of semi-major axes is smoothed, unlike triples in isolation \cite[e.g.][]{fragl19}.

\begin{figure} 
\centering
\includegraphics[scale=0.525]{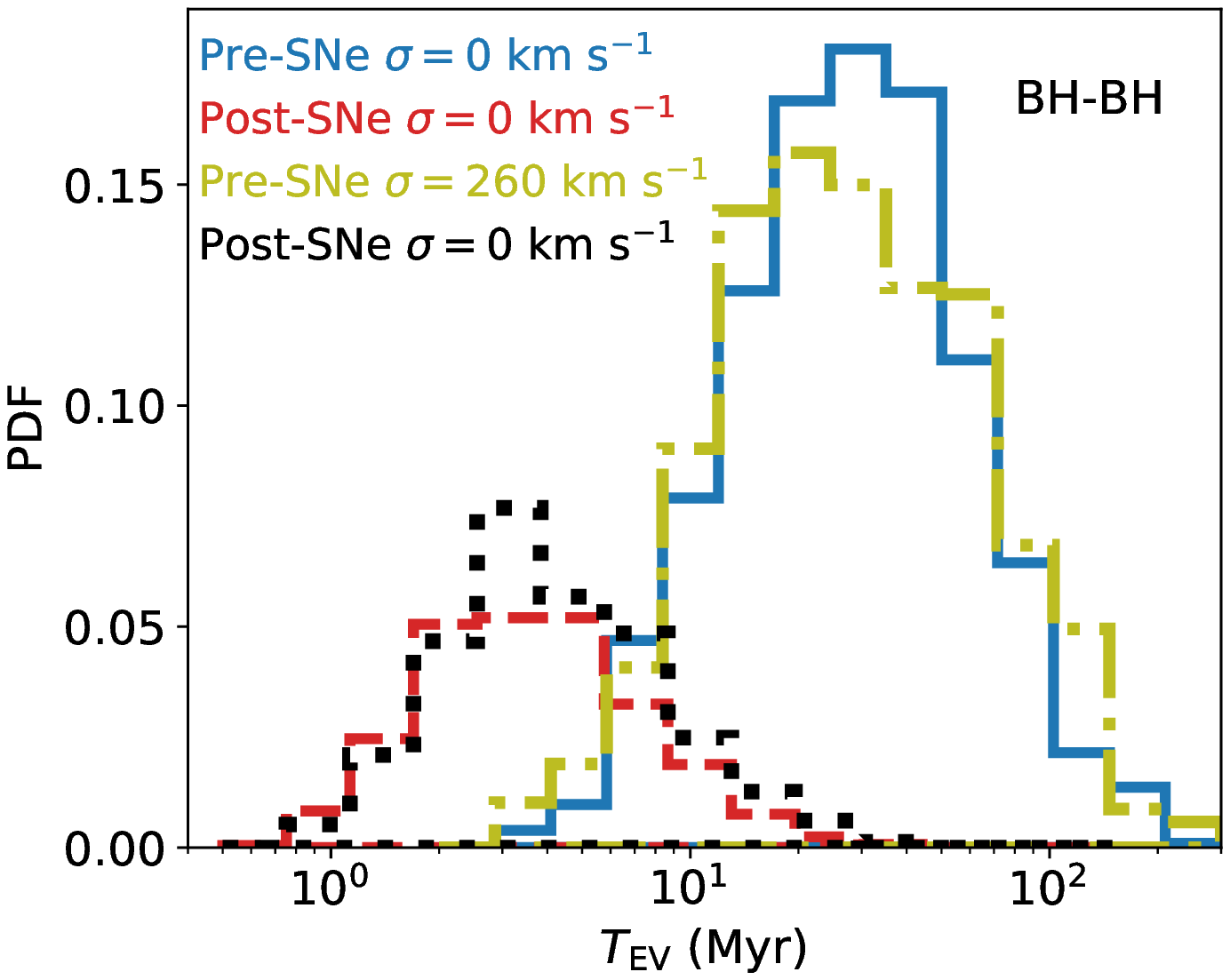}
\includegraphics[scale=0.525]{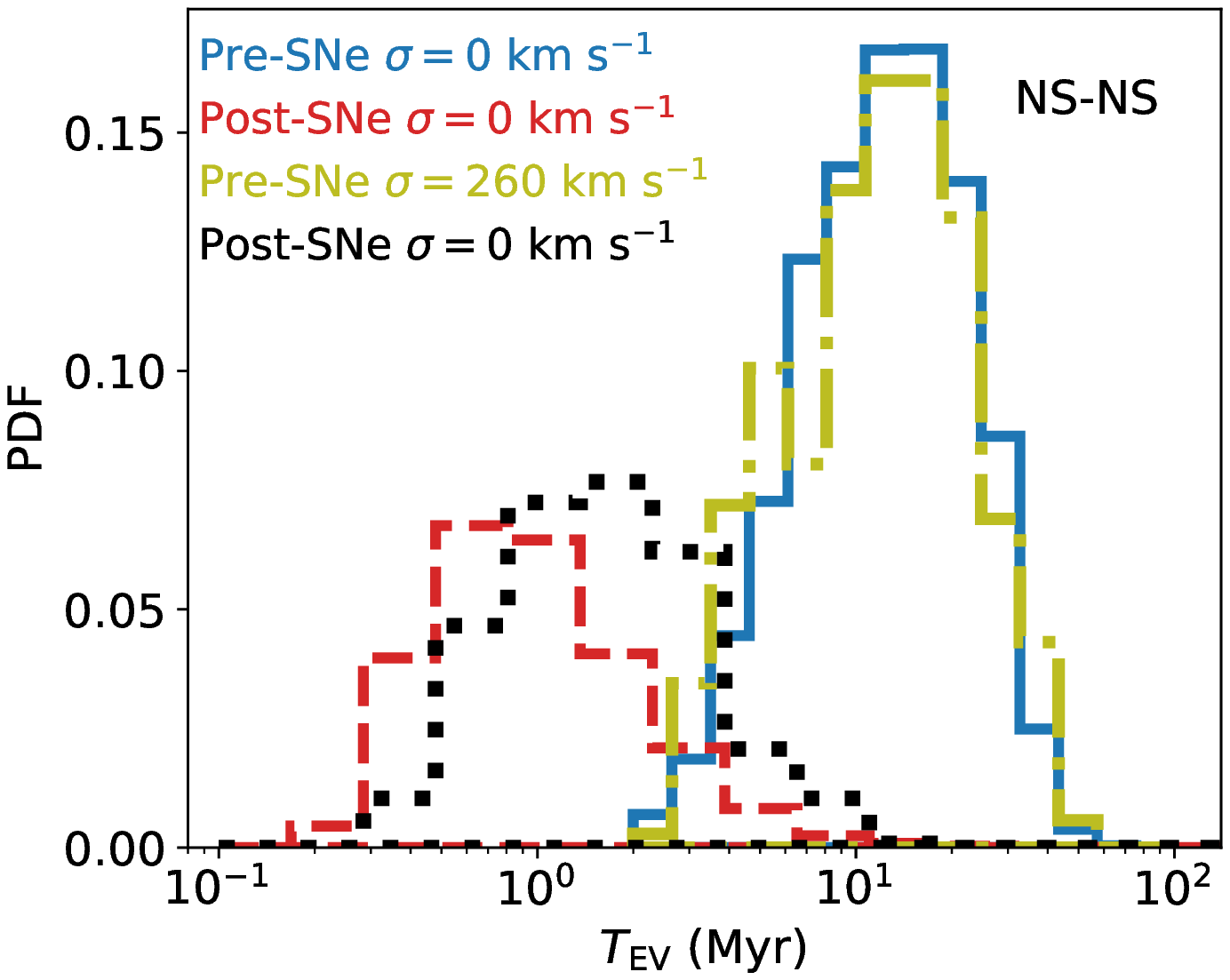}
\caption{Evaporation time of BH-BH and NS-NS binaries in a Milky Way-like galaxy pre-SNe and post-SNe, for $\sigma=0\kms$ and $\sigma=260\kms$.}
\label{fig:tev}
\end{figure}

\begin{figure} 
\centering
\includegraphics[scale=0.525]{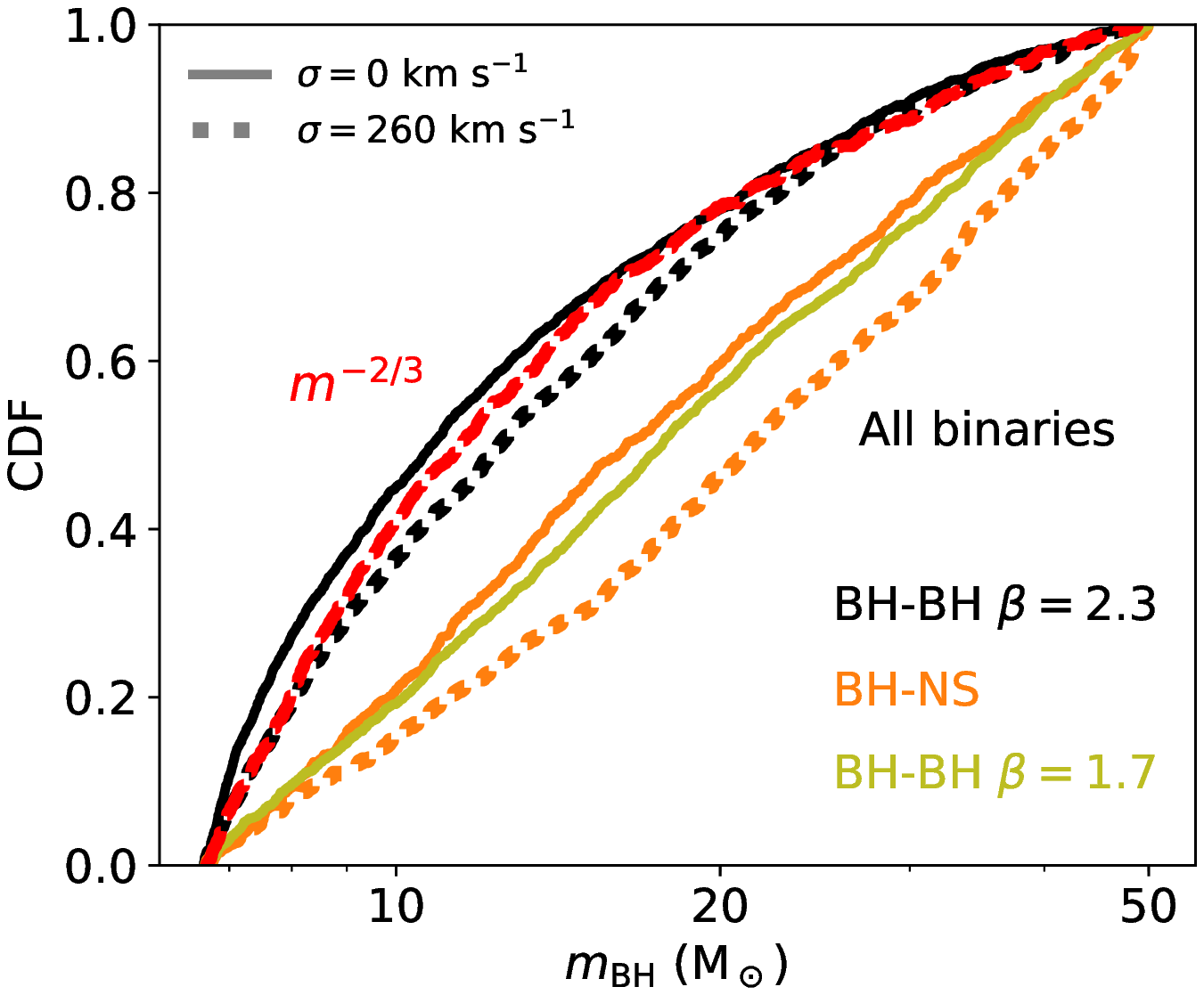}
\includegraphics[scale=0.525]{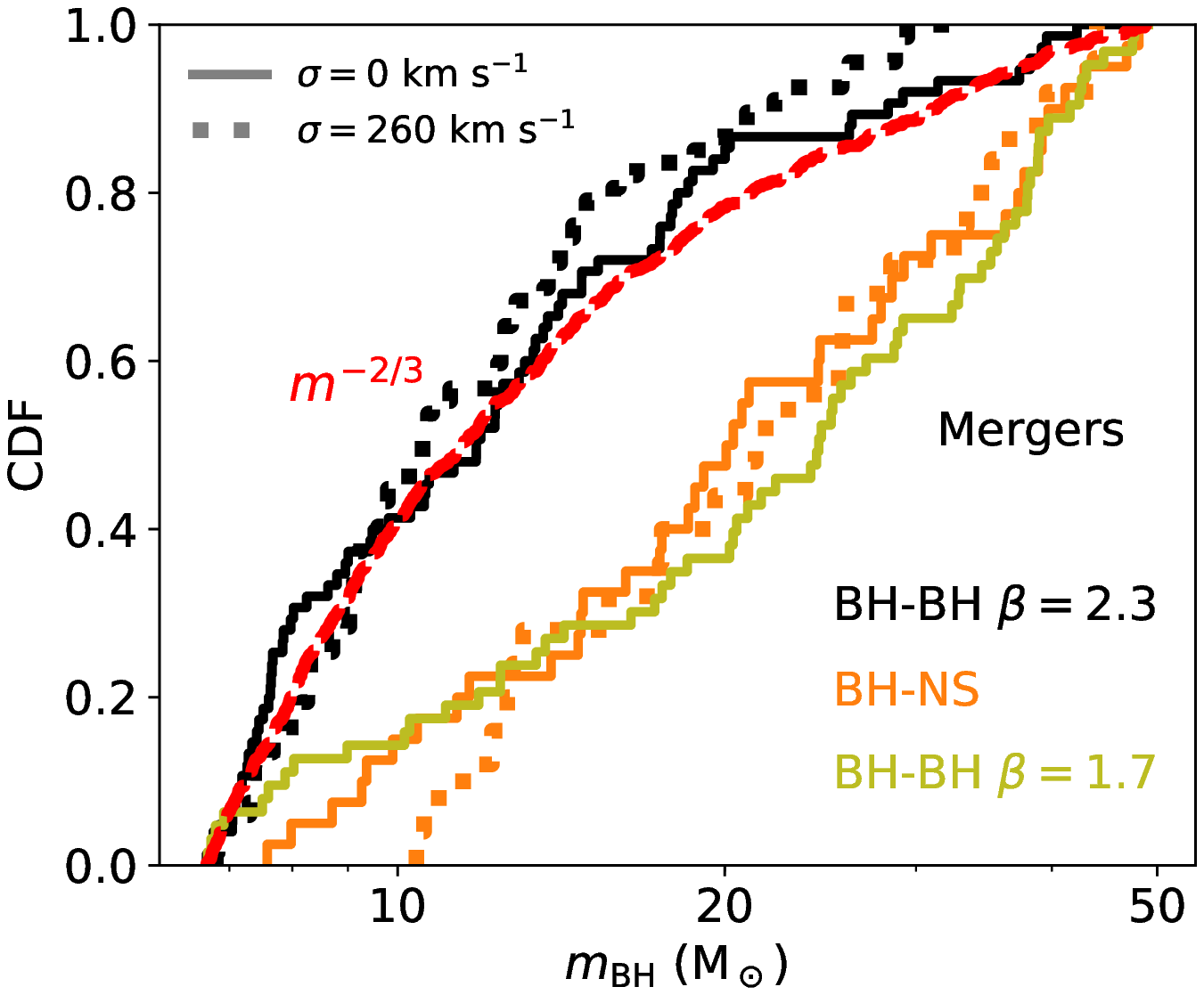}
\caption{Cumulative distribution function of BH mass of BH-BH and BH-NS binaries, for different values of $\sigma$ and $\beta$. Left panel: all binaries; right panel: binaries that lead to a merger.}
\label{fig:mass}
\end{figure}

\begin{figure} 
\centering
\includegraphics[scale=0.525]{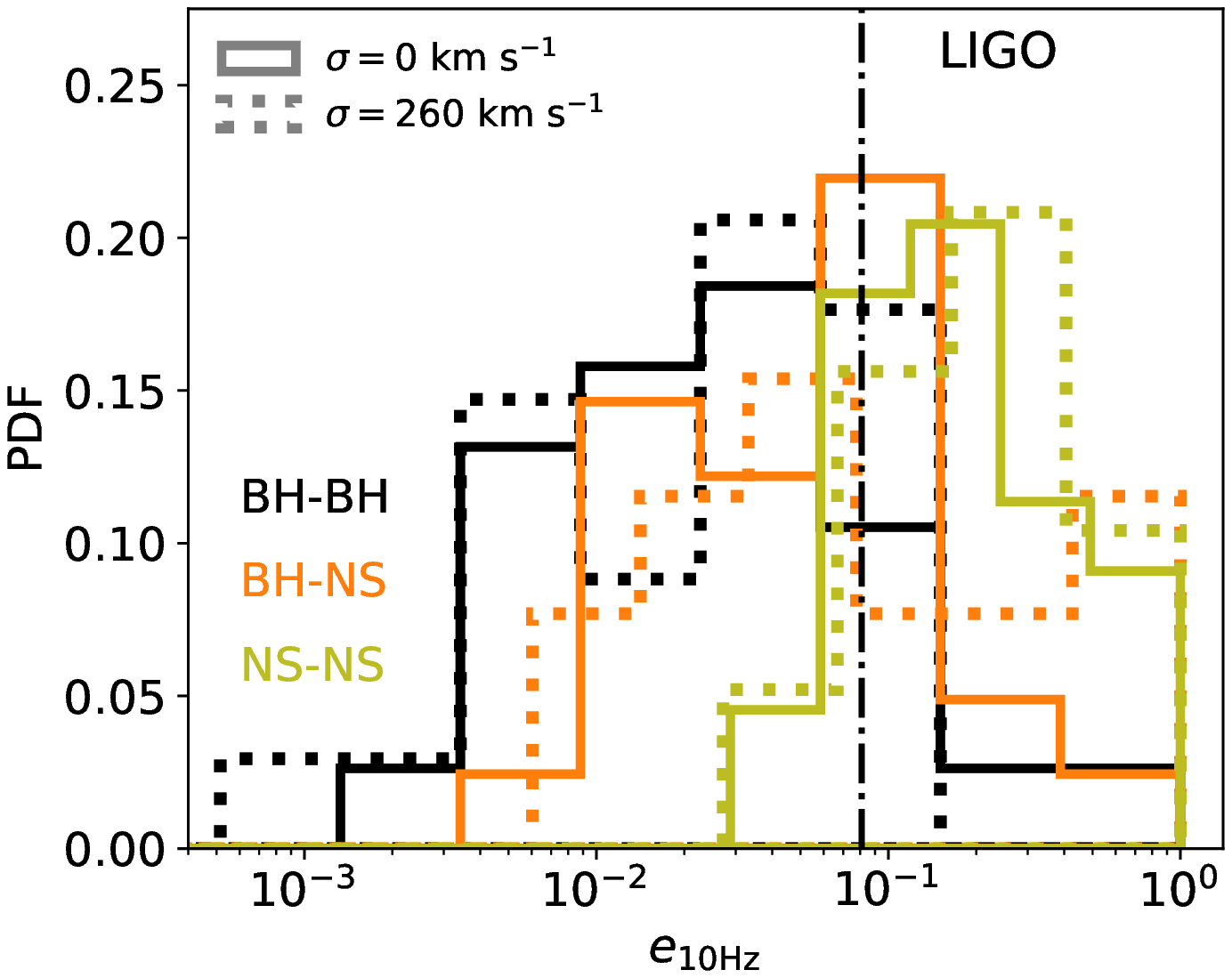}
\caption{Distribution of eccentricities at the moment the binaries enter the LIGO frequency band ($10$ Hz). The vertical line shows the minimum $e_{\rm 10Hz}=0.081$ where LIGO/VIRGO/KAGRA network may distinguish eccentric binaries from circular ones \cite{gond2019}. A noticeable fraction of binaries have a significant eccentricity in the LIGO band.}
\label{fig:ecc}
\end{figure}

\subsection{Black hole mass}

Figure~\ref{fig:mass} depicts the cumulative distribution function of BH mass of BH-BH and BH-NS binaries, for different values of $\sigma$ and $\beta$. In the top panel, we show the BH mass distribution for all binaries that are formed after the SNe take place and satisfy the stability criterion of Mardling \& Aarseth \cite{mar01}. For both BH-BH binaries and BH-NS binaries, smaller $\sigma$'s predict smaller BH masses, even though there is not a significant difference as in the case of isolated triples \cite[e.g.][]{fragl19}. We find that BH masses in BH-NS binaries are a factor of $\sim 1.5$ larger than BH masses in BH-BH binaries, in the case the initial mass function distribution $\propto m^{-2/3}$. This can be explained by the use of momentum-conservation kicks, where the larger BHs are imparted a smaller kick. As a result, binaries where the primary forms a massive BH are typically more bound and stable. These binaries have than a larger probability for surviving the second SN event, where a much larger kick is imparted to the secondary when forming a NS. In the bottom panel of Figure~\ref{fig:mass}, we plot the BH mass distribution for the binaries that lead to a merger. The CDFs of BH mass in CO binaries that lead to a merger essentially maps the initial BH mass distribution. 

\subsection{Eccentricity at LIGO band}

CO binaries merging in isolation are expected to enter the LIGO band ($10$ Hz) on almost circular orbit, i.e. with very low eccentricities ($e\sim 10^{-8}$-$10^{-7}$). On the other hand, CO binaries that merge as a result of the LK cycles due to the torque of a third companion have much higher eccentricities. In our simulation, we compute a proxy for the frequency, which we take to be the frequency that corresponds to the harmonic of the maximal GW emission \cite{wen03},
\begin{equation} 
f_{\rm GW}=\frac{\sqrt{G(m_1+m_2)}}{\pi}\frac{(1+e_{\rm in})^{1.1954}}{[a_{\rm in}(1-e_{\rm in}^2)]^{1.5}}\ .
\end{equation}
Figure~\ref{fig:ecc} shows the distribution of eccentricities at the moment the CO binaries that produce a merger enter the LIGO frequency band. We also show the minimum eccentricity, $e_{\rm 10Hz}=0.081$, where LIGO could distinguish eccentric binaries from circular ones \cite{gond2019}. A large fraction of systems have a significant eccentricity at $10$ Hz. The detectability and selection biases inherent to high-eccentric sources, as we predict in our model, are difficult to determine. A large eccentricity would also limit the effectiveness of current matched-filtering searches, which analyze data using quasi-circular templates. Recently, \cite{lvs2019} have started investigating the possible presence of eccentric sources in the first and second observing runs (O1 and O2) of Advanced LIGO and Advanced Virgo, but did not find any of them so far.

\subsection{Merger time}

\begin{figure} 
\centering
\includegraphics[scale=0.525]{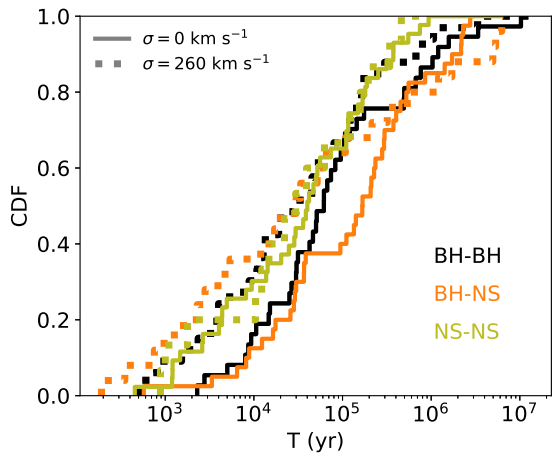}
\caption{Merger time distribution of of BH-BH, BH-NS and NS-NS binaries that merge in a Milky Way-like galaxy (see Table~\ref{tab:models}).}
\label{fig:tmerge}
\end{figure}

Figure~\ref{fig:tmerge} reports the merger time cumulative distribution functions of BH-BH, BH-NS, and NS-NS binaries that merger for all models. The cumulative function does not depend significantly on the values of $\sigma$. This can be explained by considering two competing effects, the LK cycles and the evaporation of CO binaries. Smaller kick velocities imply a larger inner semi-major axis, thus a smaller LK timescale. However, wider CO binaries orbiting an SMBH are susceptible to evaporation on shorter timescales as a result of the dynamical interactions with other stars and COs, before the LK cycles can drive them to merger. As a result, the CDF of merger times is not significantly affected by the assumed value of $\sigma$.

\subsection{Rates}

We can use the results of our $N$-body simulations combined with the observations to estimate the Co binary merger rate due to LK oscillations. The star-formation rate close to non-resolved regions around SMBHs is difficult to estimate. We use an empirical estimate based on our Galactic Centre \cite[see e.g.][]{bart09}. Approximatively $\sim 200$ O-stars (likely to later form NSs and BHs) are observed and inferred to have formed over the last $\sim 10$ Myrs in the young stellar disk (at distances $\sim 0.05$--$0.5$ pc) around the SMBH. The number of lower-mass B-stars in the same environment suggests that similar continuous star-formation may have not occurred over the last $\sim 100$ Myr. Thus, we compute an in-situ formation rate of massive stars of $\Gamma_{\rm sup}\sim 200/10^8$ yr$=2\times10^{-6}$ yr$^{-1}$. For reference, a similar rate can be computed in the case massive stars are supplied by 2-body migration \cite[ex-situ scenario, see e.g.][]{antoper12,fragrish2018}. Hence, the overall rate of mergers of CO binaries orbiting an SMBH formed in-situ is,
\begin{equation}
\Gamma(\msmbh)=n_{\rm gal} f_{\rm SMBH} \Gamma_{\rm sup} f_{\rm mb} f_{\rm mb,surv} (1-f_{\rm EV}) f_{\rm stab} f_{\rm merge}\ ,
\label{eqn:rate}
\end{equation}
where $n_{\rm gal}$ is the galaxy density, $f_{\rm SMBH}\approx 0.5$ is the fraction of galaxies containing an SMBH \cite{anto15a}, $\Gamma_{\rm sup}$ is the CO supply rate, $f_{\rm mb}$ is the fraction of massive stars in binaries, $f_{\rm mb,surv}$ is the fraction of massive stars in binaries that do not merge before collapsing to COs \cite{fragione19}, $f_{\rm EV}$ is the fraction of binaries that evaporate before collapsing to a CO binary (see Fig.~\ref{fig:tev}), $f_{\rm stab}$ is the fraction of stable binaries after the SN events (see Tab.~\ref{tab:models}), and $f_{\rm merge}$ is the fraction of mergers we find in our simulations. We neglect the weak dependence on the SMBH mass and fix $n_{\rm gal}=0.02$ Mpc$^{-3}$ \cite*{cons05}.

Plugging numbers in Eq.~\ref{eqn:rate},
\begin{eqnarray}
\Gamma(\msmbh)&=& 0.04\ \mathrm{Gpc}^{-3} \yr^{-1} \left(\frac{4\times 10^6\msun}{\msmbh} \right)^{1/4} \times\nonumber\\
&\times &\left(\frac{n_{\rm gal}}{0.02\ \mathrm{Mpc}^{-3}}\right) \left(\frac{f_{\rm SMBH}}{0.5}\right)\left(\frac{f_{\rm mb}}{0.8}\right)\times \nonumber\\
&\times & \left(\frac{f_{\rm mb,surv}}{0.8}\right) \left(\frac{1-f_{\rm EV}}{0.9}\right) \left(\frac{f_{\rm stab}}{0.1}\right) \left(\frac{f_{\rm merge}}{0.02}\right)\ ,
\label{eqn:finalrate}
\end{eqnarray}
where $f_{\rm mb}$ is the fraction of massive stars in binaries, that we normalize to the value observed in the field \cite{duq1991,sana2012}, and $f_{\rm merge}$ is the fraction of mergers we find in our simulations (see the last column in Tab.~\ref{tab:models}). We find a merger rate of $0.05$--$0.07\ \mathrm{Gpc}^{-3} \yr^{-1}$, $0.04$--$2\times 10^{-3}\ \mathrm{Gpc}^{-3} \yr^{-1}$, $9.6\times 10^{-6}$--$2.7\times 10^{-3}\ \mathrm{Gpc}^{-3} \yr^{-1}$ for BH-BH, BH-NS, NS-NS, respectively. Our predicted rate for this channel is smaller than the observed LIGO rate \cite{ligo2018}, but consistent with previous estimates \cite{antoper12,fragrish2018,hoang18}. We note that any rate estimate is highly uncertain since it relies on the specific assumptions regarding the star-formation history in galactic nuclei and the supply rate of compact objects at various distances from the SMBH, which remains poorly constrained \cite[see discussion in][]{fragrish2018}.

\section{Discussion and conclusions}
\label{sect:conc}

In this paper, we have modeled the SN processes in massive binaries orbiting an SMBH that leads to the formation of a CO binaries and studied the following dynamical evolution of by means of high-precision $N$-body simulations, including Post-Newtonian terms up to 2.5 order. We adopted different prescriptions for the natal kick velocities that are imparted by SN events and different slopes of the initial mass function. 

We have found that the majority of the BH-BH, BH-NS, and NS-NS mergers takes place when the inclination is $\sim 90^\circ$, as expected from LK quadrupole theory, and that the value of $\sigma$ affects the distribution of orbital elements of the merging CO binaries. We have showed that the larger the mean natal kick, the smaller is the semi-major axis of merging BH-BH, BH-NS, and NS-NS binaries. We have also shown that BH masses in BH-NS binaries are a factor of $\sim 1.5$ larger than BH masses in BH-BH binaries,as a consequence of momentum-conservation kicks ($\beta=2.3$). Moreover, the merger time distribution of CO binaries does not depend significantly on the assumed $\sigma$, since wider CO binaries have shorter LK timescales, but evaporate on shorter timescales as a result of the dynamical interactions with other stars and COs. We have then calculated the resulting rates as a function of the SMBH mass and found that the merger rates are $0.05$--$0.07\ \mathrm{Gpc}^{-3} \yr^{-1}$, $0.04$--$2\times 10^{-3}\ \mathrm{Gpc}^{-3} \yr^{-1}$, $9.6\times 10^{-6}$--$2.7\times 10^{-3}\ \mathrm{Gpc}^{-3} \yr^{-1}$ for BH-BH, BH-NS, NS-NS, respectively. We note that our rates refer exclusively to CO binaries that form in-situ, from the evolution of massive binaries. CO binaries that form farther out and gradually migrate in could give a larger rate, in particular for BH-BH merger \cite{antoper12,fragrish2018}. However, any rate estimate is uncertain since it relies on the specific assumptions regarding the star-formation history in galactic nuclei and the supply rate of compact objects at various distances from the SMBH, which remains poorly constrained \cite[see discussion in][]{fragrish2018}. Finally, we have also discussed that a signature of this model is the high eccentricity of merging CO binaries in the LIGO band. We note that a similar signature is typical also of some type of mergers in globular clusters \cite{sam2018,sam2018b}, binaries orbiting intermediate-mass black holes in star clusters \cite{fragbr2019}, and hierarchical triples and quadruples \cite{ant17,fragk2019}.

We caution that we have assumed very simple prescriptions to derive the BH and NS mass from the mass of their progenitors. Of course, the exact values of $\mns$ and $\mbh$ depend on the details of stellar evolution models and, in particular, on the metallicity, rotation and stellar winds. The progenitor metallicity can influence in particular the final BH mass, while stellar winds could also be important since they may change the orbital parameters of binaries before a BH or a NS is formed. This would affect the initial conditions of our BH-BH, BH-NS and NS-NS binaries and the BH mass spectrum. Moreover, depending on the metallicity and mass of the progenitors, stars may leave no remnants at all as a result of the pair-instability SN. This would lower the rate of BH-BH and BH-NS mergers we have derived by a factor $\eta<1$, where $\eta$ is the fraction of progenitors that do not undergo pair-instability SN and form BHs. This fraction can even decrease if some of the progenitors expand considerably during the giant phase, thus possibly leading to a merger before the CO binary is formed, in particular if LK cycles were to take place during this phase of the binary lifetime. These would reduce the number of systems that can later collapse to form a CO binary and, eventually, produce a CO merger \cite{shapp2013}. The picture is even more complicated if mass loss during possible episodes of Roche-lobe overflows and common evolution phases are taken into account \cite{shapp2013}. While these processes are relatively accounted for in binaries, even though with high uncertainty, they are not modeled properly in triple systems because of the interplay with the LK cycles imposed by the tertiary \cite{rosa2019,hamd2019}. Given the uncertainties on the stellar evolution in hierarchical systems, we have preferred a simple procedure to derive the final BH and NS masses. We note that our method, nevertheless, gives a maximum BH mass consistent with recent theoretical results on pulsational pair instabilities, which limit the maximum BH mass to $\sim 50\msun$ \cite{woo2016,woo2017}. 

Recently, \cite{steph2019} have investigated the dynamical and stellar evolution of binary stars in a Milky Way-like nucleus. \cite{steph2019} have started with a general binary population, similarly to \cite{steph2016M}, and focused on binaries that cross their Roche lobe and locked short-period binaries. While their method results still limited by the complicated behaviour of stellar evolution in triple systems, \cite{steph2019} have shown that some binaries may produce rejuvenated stars, G2-like objects, stripped giant stars, Type Ia SNe, cataclysmic variables, symbiotic binaries, or CO binaries. In particular, they derived a rate of $1\ \mathrm{Gpc}^{-3} \yr^{-1}$ for CO mergers, larger than our findings, but still consistent within the uncertainties.

While mergers of BH-BH binaries are more likely to be dark, except from the case the SMBH is surrounded by a gaseous disk as in active galactic nuclei. In this avenue, any BH-BH binary merging within the disk can accrete some gas, thus producing an electromagnetic (EM) counterpart \cite{bart17}. In the case of BH-NS and NS-NS mergers, the merger event could naturally produce some EM counterpart, such as short gamma-ray bursts, and can potentially provide critical information on the binary formation and the structure of NSs \cite{Pannarale2011,Foucart2018}. 

\acknowledgments
We thank Manasvi Lingam for useful comments and suggestions. GF thanks Seppo Mikkola for helpful discussions on the use of the code \textsc{archain}. GF acknowledges support from a CIERA postdoctoral fellowship at Northwestern University. IG is supported in part by Harvard University and the Institute for Theory and Computation. AL is supported in part by Harvard's Black Hole Initiative, which is founded by a JTF grant.

\bibliographystyle{JHEP}
\bibliography{refs}

\end{document}